\pdfoutput=1
\documentclass{article}

\usepackage{amsmath,amssymb}
\usepackage{amscd}

\usepackage{multirow}
\usepackage{accents}
\usepackage{hyperref}
\usepackage{color}
\usepackage[usenames,dvipsnames]{xcolor}
\usepackage{graphicx}
\usepackage{enumitem}

\usepackage{algorithm}
\usepackage{algorithmic}

\usepackage{setspace}

\usepackage[
pass,% keep layout unchanged
]{geometry}

\newgeometry{top=3.5cm, bottom=3.5cm, left=3.5cm, right=3.5cm}

\newfont{\mycrnotice}{ptmr8t at 7pt}
\newfont{\myconfname}{ptmri8t at 7pt}

\makeatletter
\renewcommand{\@oddhead}{\scriptsize A.~Drutsa et al., ``Prediction of Video Popularity in the Absence of Reliable Data  from Video Hosting Services: Utility..."\hfil}
\makeatother

\title{Prediction of Video Popularity in the \\Absence of Reliable Data  from Video Hosting Services:\\ Utility of Traces Left by Users on the Web}
\date{November 28, 2016}

\author{
	{\bf Alexey Drutsa}\\ Yandex\thanks{16, Leo Tolstoy St., Moscow, Russia (www.yandex.com)} \\ {\tt\small adrutsa@yandex.ru}
	\and 
	{\bf Gleb Gusev}\\ Yandex\footnotemark[1] \\ {\tt\small gleb57@yandex-team.ru}
	\and 
	{\bf Pavel Serdyukov}\\ Yandex\footnotemark[1] \\ {\tt\small pavser@yandex-team.ru}
}

\newcommand{\argmin}{\mathop{\mathrm{argmin}}}

\begin{document}

\maketitle

\begin{abstract}
With the growth of user-generated content, we observe the constant rise of the number of companies, such as search engines, content aggregators, etc., that operate with tremendous amounts of web content not being the services hosting it. Thus, aiming to locate the most important content and promote it to the users, they face the need of estimating the current and predicting the future content popularity.

In this paper, we approach the problem of video popularity prediction not from the side of a video hosting service, as done in all previous studies, but from the side of an operating company, which provides a popular video search service that aggregates content from different video hosting websites.
We investigate video popularity  prediction based on features from three primary sources available for a typical operating company: first, the content hosting provider may deliver its data via its API; second, the operating company makes use of its own search and browsing logs; third, the company crawls information about embeds of a video and links to a video page from publicly available resources on the Web.
We show that video popularity prediction based on the embed and link data coupled with the internal search and browsing data significantly improves video popularity prediction based only on the data provided by the video hosting and can even adequately replace the API data in the cases when it is partly or completely unavailable.
\end{abstract}


\section{Introduction}

With the stunning growth of user-generated content, we observe the constant rise of the number of companies that operate with web content not being services hosting it. In this respect, we can distinguish two types of companies. The first ones are the organizations that provide a hosting service for user content (\emph{hosting providers}, or \emph{HPs}). For instance, they are video hostings like Youtube, music sharing services like Soundcloud, etc.  The second ones (\emph{operating companies}, or \emph{OCs}) are the organizations that operate with user content which is hosted externally at HPs or other OCs. Examples of operating companies are  web search engine companies (e.g., Google, Bing), content aggregators (e.g., Digg, Reddit), content recommendation systems (e.g., StumbleUpon, Pinterest), etc. Of course, one company may act both as HP and OC. For example, large social networks like Facebook and Twitter store billions of user messages and, at the same time, they provide the ability to embed external videos and images directly into the messages.

\begin{figure}
	\centering
	\includegraphics[width=\textwidth]{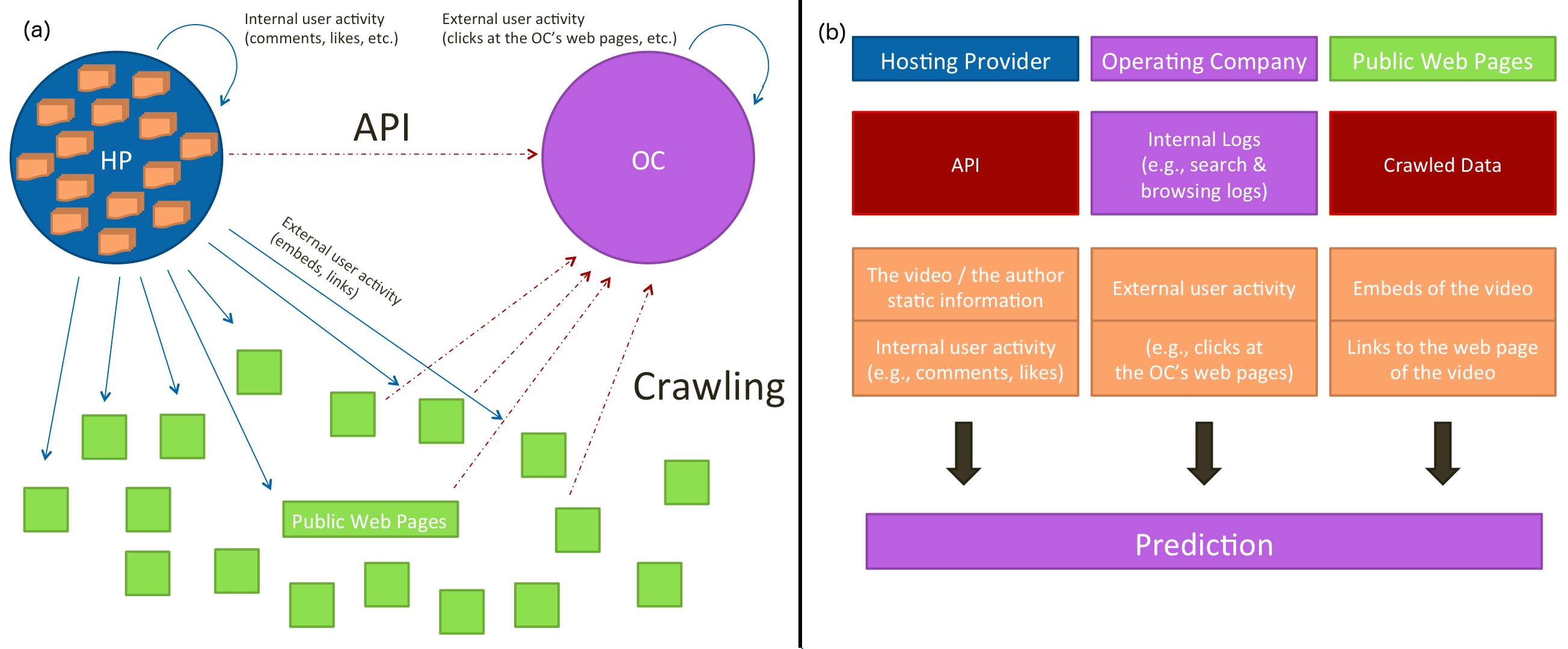}
	
	\caption{(a) A Hosting Provider (HP), an Operating Company (OC), and several public web pages that accommodate user activity about content items from the HP; (b)  all available sources of evidence about current or future video popularity that could be split into the following three groups: API data provided by the HP, internal logs of the OP, and crawled data from publicly available web pages.}
	
	\label{img_descr}
\end{figure}

Since  operating companies usually deal with tremendous amounts of external content, the challenge of estimating the current and the future popularity (e.g., the number of views, the number of comments received, etc.) of the content is inevitable for them. It is considered that the predicted current and future values of content popularity can serve as strong features for content ranking and content analysis  problems in general \cite{2010-IC-Goncalves,2012-ASONAM-Tatar,2012-WSDM-Yin,2013-WSDM-Ahmed}. So, a high quality popularity  prediction  is a vitally important component of any OC, which strongly influences the usefulness of the service to its end users and, consequently, the company's profits.

In some situations, the popularity of the content is disclosed by the hosting provider through an application programming interface (API); in other circumstances it could not be retrieved from the HP at all (the API could be simply absent, as, for instance, for the video hosting services coub.com and break.com). Meanwhile, even if the API provides the information on popularity,  the API could be periodically or permanently unavailable, or could set a limit on the number of allowed requests per time period, which can be insufficient for the OC's needs. Besides, the provided data could be noisy and could be delivered with a delay as we demonstrate further with an example. In our work, we propose a methodology that could be used to compensate for the above limitations faced by operating companies.

In the current paper, we solve the popularity prediction problem and we restrict our investigation to the needs of a company which provides a popular video search service and aggregates content from different video hosting websites. Moreover, without a loss of generality, we will only consider the video data served by the video hosting website Youtube.

On the one hand, it is well known that the video popularity (the total number of views during a considered period) in vast majority of cases depends on or, at least, highly correlated with its popularity in the first days of its existence \cite{2010-CACM-Szabo,2011-WSDM-Figueiredo,2013-WSDM-Pinto,2013-JIIS-Broxton}. Thus, the long-term popularity could be efficiently predicted using information about popularity dynamics in the first weeks of the video existence. 
On the other hand, it is known that Youtube can freeze the views count in the first days of video existence. It is officially confirmed by Youtube representatives\footnote{``We want to make sure that all views are validated so during this process the views update less frequently and might occasionally freeze above 300 views to assure quality view count. This is the normal operation in YouTube videos." stated on the official Youtube site \url{support.google.com/youtube/troubleshooter/2991876}} and is also frequently observed in the scope of the video search service of the popular search engine company under study.
Thus, we face the problem of popularity prediction of a new video in the first days of its existence without knowing its previous popularity, because  the information from Youtube is not always available.

In this paper, we propose to consider the case when an operating company decides not to stay completely dependent from video hosting providers and relies not on a single, but on all available sources of evidence about current or future video popularity that could be split into the following three groups (see Fig.~\ref{img_descr}):
\begin{itemize}
	\item the data provided by the content hosting provider, generally, via its API or its publicly available web pages (API data);
	\item the internal data of the content operating company, generally, its user access data stored in logs (Log data);
	\item the publicly available resources of the Web, where the content could leave its traces (Web data).
\end{itemize}

The future popularity of objects of different kinds and videos in particular has been investigated and predicted from the point of view and using the data from content hosting providers only \cite{2008-PNAS-Crane,2010-CACM-Szabo,2010-WWW-Lerman,2010-AIR-Tsagkias,2010-IC-Goncalves,2010-NOSSDAV-Lai,2011-WWW-Hong,2012-KDD-Borghol,2012-CIKM-Kupavskii,2012-KDD-Kim,2012-WWW-Radinsky,2013-WSDM-Pinto,2013-WSDM-Figueiredo,2013-WSDM-Ahmed,2013-JIIS-Broxton,2013-WSDM-Pinto,ding2015video,fontanini2016web,li2016characterizing} and using internal data of social networks, such as Facebook and Twitter \cite{2013-CIKM-Li,2013-CIDM-Soysa,2013-ChSIP-Soysa,2014-WSDM-Abisheva}.
{
\sloppy 

}

A comprehensive overview of various research questions, methodologies
and approaches in the field of prediction of web content (not only video) popularity can be
found in \cite{2014-SNAM-Tatar}.
To the best of our knowledge,  no existing study investigated the utility of publicly available traces of videos on web pages for the task of video popularity prediction. In the current paper, embeds of videos and links to video pages on publicly available resources of the Web  are considered and used as features to predict video popularity. We also conduct a detailed investigation of the features  extracted from all three groups of sources, where we use (the first to our knowledge) web search logs and browsing logs collected by Yandex (www.yandex.com) as internal data. The investigation of both sources and a series of experiments demonstrating to what extent both sources are capable to supplement or, more importantly, replace the API data in the task of video popularity prediction, \emph{represent the first major contribution of this study}.

We are also the first who thoroughly investigated the prediction of popularity of a new video in the first days of its existence and the first who addressed the problem of the \emph{current} video popularity prediction, in the absence of the information on popularity from the video hosting service. \emph{We regard this as the second major contribution of this study.} 

The rest of the paper is organized as follows. In Section~\ref{sec_RelWork}, the related work is presented. In Section~\ref{sec_Notations}, we introduce our notations and the framework. In Section~\ref{sec_ProblemStat}, prediction task is described in detail and the research questions are stated. We describe our data sets in Section~\ref{sec_DataSets}, list the set of features in Section~\ref{sec_Features}, and describe the models used for the popularity prediction in Section~\ref{sec_Models}. In Sections~\ref{sec_ExpsSetup} and~\ref{sec_ExpsRes}, we present our experiments and discuss their results. In Section~\ref{sec_Conclusions}, the study's conclusions and future work are presented.

\section{Related work}
\label{sec_RelWork}
We compare our research with other studies in three aspects. The first one relates to the video popularity analysis, the second one concerns the  future video popularity prediction, and the last aspect refers to popularity prediction studies in general.

\subsection{Video popularity analysis}
The video hosting content and its popularity were widely investigated in recent years.
In one of the most cited studies on the topic \cite{2008-PNAS-Crane}, researchers focused on the analysis of video views dynamics decay after its peak.
Both \cite{2010-CACM-Szabo} and \cite{2011-WSDM-Figueiredo} examined how quickly a video can become popular. They found that the most popular and viral videos receive the major part of their views in the first weeks of their existence.  
Another study \cite{2013-JIIS-Broxton} also found that, in general, viral videos are mostly viewed in the first week after their upload.  
All the mentioned studies confirm the importance of studying and predicting video popularity in the first days of its existence.

\subsection{Future video popularity prediction}
The authors of \cite{2010-CACM-Szabo} studied Youtube content popularity and established linear dependence between the logarithmic views counts measured at the 10-th day and at the 30-th day after the day of the video upload. 
The authors of \cite{2013-WSDM-Ahmed} used the same data, but proposed to predict the future popularity  by using  a model of content propagation through an implicit graph induced by 
patterns of temporal evolution of video popularity. 
Prediction of  the popularity peak day of a video was  studied in \cite{2014-ICMR-Jiang}.
All described approaches are not applicable in our case, because, in order to predict future popularity, they exploit currently or/and previously observed popularity that is not always available by the statement of our problem.

The research described in \cite{2013-CIKM-Li} was devoted to prediction of future video popularity in terms of the shares of a video in online social networks like Facebook.  The analogous work was made in \cite{2013-CIDM-Soysa,2013-ChSIP-Soysa}: they collected the share data not being inside the social network company, but by receiving them from end users.
Similar study of sharing behavior were carried out for Twitter \cite{2014-WSDM-Abisheva}. 
Further studies concentrated on more sophisticated models and features like sentiments extracted from video frames and user comments~\cite{ding2015video,fontanini2016web}.
The described methods could not be used in our work because they use either the data that is not publicly available for third parties, or the APIs of those social platforms. It means that a search engine which would decide to rely on that data would need to at least use the APIs of those services, while the goal of this study is to show to what extent an operating company can be independent from any APIs, even from seemingly more important APIs of video hosting services.

In Tables~\ref{FactorTable_API} and~\ref{FactorTable_LOGWEB} we listed all features that we used to predict videos' popularity. The contents of the table will be discussed in detail further in Section~\ref{sec_Features}, but, at the moment, we are interested in the column named ``Description (previously used or new)", where we pointed out for each feature whether it has been used elsewhere in the literature, or it is a novel one.  A reference is written in \emph{italic style}, if the corresponding feature has been analyzed for a different purpose or has been used as a feature to predict the popularity of any  object other than a video, but has \emph{not} been used to predict videos' popularity. Otherwise the reference is written in normal style. 

The most comprehensive study of the features that could be retrieved via the Youtube API has been conducted in \cite{2012-KDD-Borghol} devoted to the content agnostic factors. Some of the features were also examined in \cite{2013-WSDM-Figueiredo,li2016characterizing}. Most of them are in  Table~\ref{FactorTable_API}, but not all: the researchers have used the number of keywords assigned to a video, the number of times the video was ``favourited",  and the best quality the video is available in. All these features are still provided by Youtube API for old videos, but they seem to be deprecated for new videos: their values for all videos that were uploaded since 2013 are constant (zero, for instance). The reason is that Youtube does not allow its users to assign keywords and favorite videos anymore. In addition, the results of \cite{2012-KDD-Borghol} state that these features have no significant correlation with video popularity and the  ``favourited" feature has  a strong correlation with other user feedback features (numbers of comments, ratings, likes). Therefore, we do not use them in our analysis. Nevertheless, these circumstances serve as an additional confirmation of any API inconsistency  and indicate the high chance that the provided data could become obsolete or unavailable at any time.

Note that all above-mentioned studies and their corresponding features were used to predict future video popularity only, while our study also focuses on the prediction of current video popularity. Therefore, we reproduced all the features available through Youtube API and used them for our baseline models.

\subsection{Popularity prediction of other objects}
Prediction of web content (not only video) popularity is a well known and widely investigated problem \cite{2014-SNAM-Tatar}. Prediction methods similar to the ones described in the previous subsection were applied to predict popularity of web pages in general \cite{2010-CACM-Szabo,2010-WWW-Lerman,2012-EPJDS-Hogg} and for popularity of news measured in comments count \cite{2010-AIR-Tsagkias}. 
The popularity of tweets in terms of the number of retweets and shows were studied in \cite{2011-WWW-Hong,2012-CIKM-Kupavskii,2013-CWSM-Kupavskii}. 

Some studies consider popularity prediction models that utilize data in non-aggregated form (like news articles \cite{2010-ICDM-Yang} and user comments \cite{2014-SIGIR-He}).
The authors of \cite{2010-ICDM-Yang} introduced the linear influence model in order to predict popularity of hashtags over Twitter network and popularity of memes over news articles and blog posts in terms of affected nodes of an implicit network. In our work, we use this model in order to take into account hosts with embeds and links to videos in non-aggregated form.

The usefulness of content popularity prediction for search engines was discussed in  \cite{2012-ASONAM-Tatar,2012-WSDM-Yin,2014-SNAM-Tatar} where the prediction quality was estimated with ranking metrics for popularity of news articles and published jokes.
Hence, in our study, we also evaluate the performance of our predictors by means of NDCG (one of the most popular ranking metrics~\cite{2002-TOIS-Jarvelin}).
{
\sloppy
	
}

\section{Notations and framework}
\label{sec_Notations}
Let $[0,T),\: T > 0,$ be a known time interval, and $\tau$ be a fixed time step (e.g., one day in our experiments). Then $\tau$ induces the finite time grid $\overline{\mathbb{T}} = \{t_m\}_{m = 0}^{M} \subset [0,T)$, where $t_m = m\tau, m = 0,..,M$. The mesh without the starting point (e.g., the starting day) is denoted by $\mathbb{T} = \overline{\mathbb{T}} \setminus \{0\}$. 

Let $\mathcal{V}$ be a set of objects. Each object $v \in \mathcal{V}$ is created at some time moment $t_\circ(v)$ and exists at all times $t \ge t_\circ(v)$. From here on in the paper \emph{we assume} that for each object $v\in \mathcal{V}$, there is chosen the object-specific time scale measured in days and centered in such a way that $t_\circ(v)\in [0,\tau)$, i.e. the object is created on the starting day of the scale.
For instance, a video published at 18:00 7th May has $t_\circ(v)=0.75\tau$ for $\tau = 1$ day. These object-focused timelines allow to consider examples of objects created at different days in the scope of the same prediction tasks we described further.

Each object can be represented by the features from some set $\mathbf{\Phi}$ during its life (for instance, video duration, number of comments, etc.). Each feature $\varphi\in\mathbf{\Phi}$ takes its values in a set $\mathbb{D}_\varphi$. Generally, $\mathbb{D}_\varphi$ is the set of real numbers $\mathbb{R}$ (e.g., for average rating), the set of integers $\mathbb{Z}$ (e.g., for number of views),  a finite set (e.g., for video category),  their Cartesian product, or their subset. 
Besides, features could take different values at different time moments $\overline{\mathbb{T}}$, being dynamic in nature. Although, there are static features  that are known either at or before the object creation time (such as video upload hour). Thus,  each  feature $\varphi\in\mathbf{\Phi}$ is formally a map from object set $\mathcal{V}$ \emph{and the time mesh} $\overline{\mathbb{T}}$  to the feature value set $\mathbb{D}_\varphi$, that is
$
\varphi : \mathcal{V}\times\overline{\mathbb{T}} \rightarrow \mathbb{D}_\varphi, \: \varphi \in \mathbf{\Phi}.
$

It is common that some data are known for the observer and some other are not. Usually, the observer wants to use some part of the known data to estimate or predict the unknown data. The set of features used to predict unknown data is denoted  by  $\mathbf{\Psi}\subset\mathbf{\Phi}$. The features that the observer wants to estimate or predict are referred to as \emph{targets} and their set is denoted by $\mathbf{\Theta}\subset\mathbf{\Phi}$.

Let the values of features $\mathbf{\Psi}$ be known for the observer at the time moments $\overline{\mathbb{T}} \cap [0,t_c]$. Then the time moment $t_c \in \overline{\mathbb{T}}$ is referred to as \emph{the current time moment}. Consider that the observer has to estimate or predict the value of a target  $\theta\in\mathbf{\Theta}$ at the time moment $t_t \in \overline{\mathbb{T}}$. Then the time moment $t_t$ is called \emph{the target time moment}.
Note that, if $t_c = t_t$, then we are dealing with \emph{current prediction}; if $t_c < t_t$, then we are dealing with \emph{future prediction}\footnote{The current prediction task usually rises when the values of the target $\theta$ could not be retrieved (e.g., the current number of views from Youtube API), while the future prediction task could be stated for any feature.}.

Thus, for a fixed target $\theta\in\mathbf{\Theta}$, a fixed feature set $\mathbf{\Psi}'\subset\mathbf{\Psi}$, fixed current and target time moments $t_c, t_t \in \overline{\mathbb{T}}$, and a fixed prediction model $\mathbf{m}$ there is  a class of functions $\mathfrak{P}_{\mathbf{m}}(\mathbf{\Psi}', \theta) $  that predict the
target $\theta$ based on the features $\mathbf{\Psi}'$. Then,
the problem of prediction in terms of machine learning could be stated as follows. Given a training set of examples $\mathcal{V}'\subset\mathcal{V}$ and a \emph{prediction performance metric} $\rho_\theta$ on the function space $\{\mathcal{V}'\rightarrow \mathbb{D}_\theta\}$, one should find the optimal predictor $P_{\mathbf{m},\mathrm{opt}(\mathbf{\Psi}', t_c; \theta, t_t)}$, namely 
\begin{equation}
\label{eq:minimizatioTask}
P_{\mathbf{m},\mathrm{opt}(\mathbf{\Psi}', t_c; \theta, t_t)} =   
\argmin_{P\in\mathfrak{P}_{\mathbf{m}}(\mathbf{\Psi}', \theta)}
  \rho_\theta\Big(P\big|_{t = t_c}, \theta\big|_{t = t_t}\Big) .
\end{equation}

\section{Problem statement}
\label{sec_ProblemStat}
In this section we present the research questions that we answer in our study, and we specify the prediction task by defining current and target days, target values that we aim to predict and the metrics that we optimize on the training data and measure on the test data.

\subsection{Research Questions}
The main goal of our study is to identify the benefit of non-API data for the task of prediction of video popularity. Thereupon, we translate this objective into the following  research questions:
\begin{itemize}
 
	\item {\bf[RQ1]} Could the prediction quality be improved by using the Web and the Log data in addition to the data from the video hosting?
 
	\item {\bf[RQ2]} Could the Web and the Log data be effectively used in the case of absence of any reliable hosting data?
 
	\item {\bf[RQ3]} Could the Web and the Log data replace a part of the hosting service data without a significant loss in prediction quality?
 
\end{itemize}

\subsection{Specification of prediction task}
In the paper, in accordance with the notations introduced in Section~\ref{sec_Notations}, our investigation object is a video uploaded to Youtube video hosting, and, thus, the set $\mathcal{V}$ is a set of such videos. The fixed time interval $\tau$ is equal to \emph{one day}, and from here on in the paper we assume, for notation simplicity, that a unit of a time line is one day, i.e., $\tau=1$. 

We investigate video popularity in terms of the number of views received by the video. This target could be defined in different ways, but we will consider  the two most practical definitions of the target:
\begin{itemize}
 
	\item \emph{cumulative popularity} $\mathtt{Views[c]}$: the total number of views received since the video creation, that is, in the time period $[0, t)$;
 
	\item \emph{daily popularity} $\mathtt{Views[d]}$: the total number of views received during the last day, that is, in the time period $[t - 1, t)$.
 
\end{itemize}
In addition, both cumulative and daily views counts are logarithmically transformed\footnote{From here on in the paper we use $\log(x) \stackrel{def}{=} \log_2(x + 1)$, and, for each $\mathtt{\varphi}\in\mathbf{\Phi}$, we denote the logarithmic feature by $\mathtt{log(\varphi)}$.}  in order to better catch the differences between values of different magnitudes. Thus, the complete set of targets in our study is 
\begin{equation}
\label{eq:Theta_def}
\mathbf{\Theta} = \{\mathtt{Views[c]}, \mathtt{Views[d]}, \mathtt{log(Views[c])}, \mathtt{log(Views[d])}\}.
\end{equation}

In our work, we mainly focus on the task of the current popularity prediction ($t_c=t_t$), given that the API does not provide us with this particular information (e.g., by delaying it) or given that the API is entirely or partly unavailable. However, we apply the framework to the prediction of the future popularity as well. For instance, for the 1-3 days forecast, $t_t=t_c + \delta,\:\delta\in\{1,2,3\}$.
We will consider target days of the two first weeks of video existence, that is, $t_t\in\mathbb{T}^* \stackrel{def}{=} \{1,..,14\}$. 
The prediction in these target days is much more complicated than the prediction in the more distant time moments, because
for large values of $t_c, t_t$ (10 day $\rightarrow$ 30 day),
the linear dependence 
of $\mathtt{log(Views[c])}(t_t)$ and $\mathtt{log(Views[c])}(t_c)$ was established \cite{2010-CACM-Szabo,2013-WSDM-Pinto}, which makes the prediction straightforward.

We use \emph{the root mean squared error} (\emph{RMSE}) as the minimization metric. RMSE for given maps $P_1, P_2: \mathcal{V}' \rightarrow \mathbb{D}_\theta$ is defined  on a set $\mathcal{V}' \subset \mathcal{V}$ by $(\mathrm{RMSE}(P_1, P_2))^2 = |\mathcal{V}'|^{-1}\sum_{v\in\mathcal{V}'} (P_1(v) - P_2(v))^2$. The values of RMSE of the best predictor $P_{\mathbf{m},\mathrm{opt}( \mathbf{\Psi}, t_c; \theta, t_t)}$ on the test set are denoted by
\begin{equation}
\label{eq:RMSE_def_on_opt}
\mathrm{RMSE}(\mathbf{\Psi}, t_c; \theta, t_t) = \mathrm{RMSE}(P_{\mathbf{m},\mathrm{opt}( \mathbf{\Psi}, t_c; \theta, t_t)} , \theta).
\end{equation}
We also analyze the performance of different predictors using a task-specific variant of ranking quality measure NDCG described in Section~\ref{subsec_PerfMeas}.

So, to finish the specification of prediction task  one needs to specify the set of all investigated features $\mathbf{\Psi}$ and the used model $\mathbf{m}$. They will be specified in Sections~\ref{sec_Features} and~\ref{sec_Models} following the data sets description given in the next section.

\section{Data sets}
\label{sec_DataSets}
As we previously mentioned, we analyze three major data sources available for a typical OC. In our case, we used the following datasets: 
\begin{enumerate}[label=(\alph*)]
\item content hosting provider (in our case, Youtube); 
\item the internal logs of Yandex;
\item the traces that are left by the content on the Web. 
\end{enumerate}
	
The data from the hosting provider was split into two parts. The first part is the data about each video and its author provided through the Youtube API. The second part is the historical information on a video that is shown on the tab ``Statistics" of the video web page  (cumulative/daily views/shares/subscribers counts and total watch time)\footnote{The data are provided with a delay of 2-3 days. The authors of \cite{2013-WSDM-Figueiredo} and \cite{2011-WSDM-Figueiredo} used the same methodology to extract views history.}. About 25\% of videos are private or have no publicly available historical information.  We collected only those videos for which the access was publicly available. 

The data from Yandex could be also split into different parts corresponding to the logs of different services/products of the company. In our work, we investigate the logs of two services/products: the search service and the browser. The third major data source is provided by the web crawler of Yandex that supplies the two types of data about created videos: embeds of the videos and links to the videos' pages.

So, our  methodology for collecting data is as follows. First, we define a \emph{harvest time period}. Then, at each day in this period we go to the database of the crawler (which also regularly crawls Youtube's RSS feeds of the most popular and new videos\footnote{\url{http://www.youtube.com/t/rss_feeds}}), and retrieve all available Youtube videos that are created in the defined time period. At the end of each day, we retrieve the data for these videos available via Youtube API. Thus, we obtain API information for each video for each day  from the defined period. 

When the harvest period is over,  we retrieve the history of actual view counts per day for each known video. After that, we utilize  search and browsing logs of Yandex and retrieve all available information about the known videos. Finally, we retrieve embeds of the videos and links to the videos from the crawler database.

\begin{figure}
	\centering
	\includegraphics[width=\textwidth]{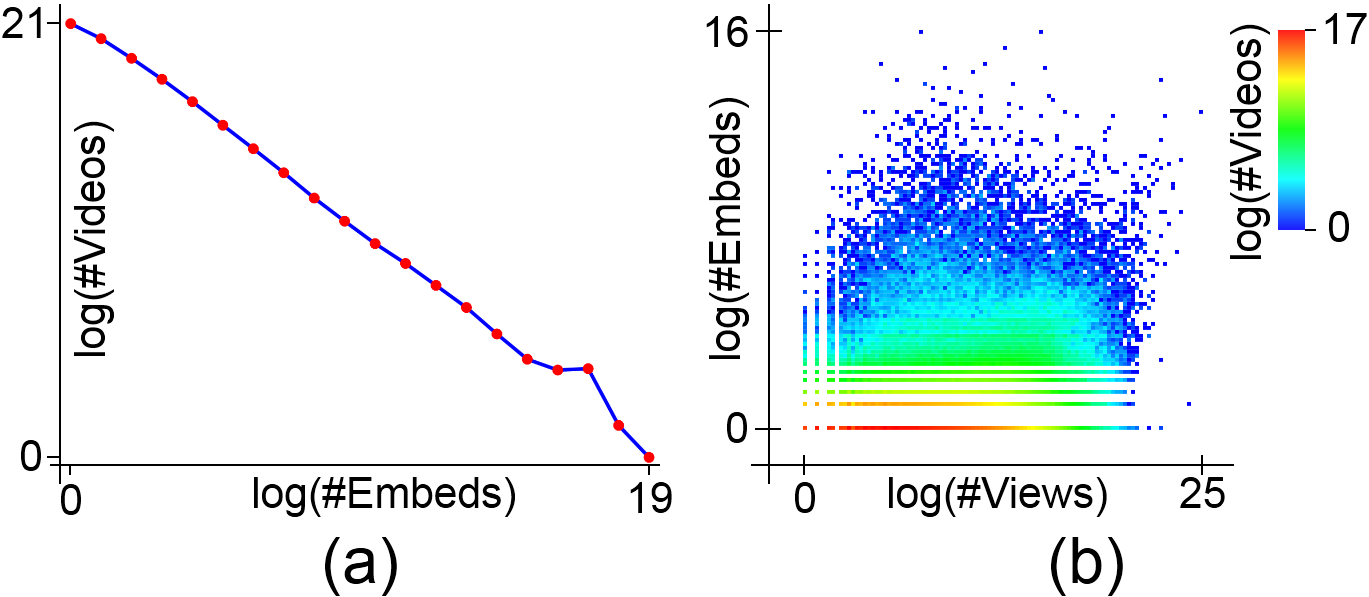}
 
	\caption{(a) The distribution of videos by the number of embeds obeys the power law (Dataset\#2); (b) The joint distribution (heat map) of videos by the number of views (axis $X$) and the number of embeds (axis $Y$) for Dataset\#1.}
 
	\label{img_embDistrib}
\end{figure}

{\bf Dataset\#1.}
For the first dataset we selected the harvest time period from 23rd December, 2013 to 18th January, 2014 (27 days period). In that period we collected 
more than $2.4 \cdot 10^6$ videos. The dataset has more than $5.3 \cdot 10^6$ embeds of the videos (more than $2.4 \cdot 10^5$ videos have at least one embed). The joint distribution of videos by the number of views and the number of embeds is presented in Fig.~\ref{img_embDistrib} (b).

{\bf Dataset\#2.}
For the second dataset we selected the harvest time period from the 1st March, 2013  to the 31st May, 2013 (92 days period). Here we did not retrieve the API data for each day unlike for the previous dataset, because  Dataset\#2 is not used in the experiments with the API data. We collected it in order to train the linear influence model (see Section~\ref{sec_Models}), and hence we also collected all embeds and links  of the discovered videos that were published in the first 120 days after each video's creation day. Overall, we collected more than $8.5 \cdot 10^6$ videos. The data set has more than $58 \cdot 10^6$ embeds of the videos (more than $3.3 \cdot 10^6$ videos has at least one embed). The distribution of videos by the number of embeds obeys the power law, see Fig.~\ref{img_embDistrib} (a).

\section{Features}
\label{sec_Features}

\begin{table*}
	\centering
	\caption{List of all features from {\tt API} that are used to predict video popularity.
		\label{FactorTable_API}}{%	
		\begin{tabular}{|l|l||p{1.3cm}|c|p{6.8cm}|p{0.9cm}|} 
			\hline
			\multicolumn{2}{|c||} {\bf Feature set }  & \centering  {\bf Name } & {\bf Values } & \multirow{2}{*}{\bf Description  (previously used  or new)} & \multirow{2}{*}{\bf Modes}\\ 
			\multicolumn{2}{|c||} {\bf $\mathbf{\Psi}$}  & \centering  {\bf $\psi$} & {\bf $\mathbb{D}_\psi$ } &  &    \\ 
			\hline
			\hline
			\multirow{14}{*}{$\mathtt{API_S}$} & & \multicolumn{4}{|l|}{ \bf\emph{ 1. Static features from the video hosting service API} } \\
			 \cline{2-6}
			
			 \cline{2-6}
			& \multirow{7}{*}{$\mathtt{API_{Sv}}$} & \multicolumn{4}{|l|}{\emph{  1.a. Static features  about the video from video hosting service API}} \\
			 \cline{3-6}
			& & {\tt Dur} & $\mathbb{N}$  & { \scriptsize Video duration in seconds (\cite{2013-WSDM-Figueiredo,li2016characterizing})} & {\small\tt n/l}  \\ 
			 \cline{3-6}
			& &  {\tt Cat} & $(\mathbb{Z}_2)^c$  & { \scriptsize Video category, where $c\in\mathbb{N}$ is the number of different categories (\cite{2013-WSDM-Figueiredo,li2016characterizing},  \emph{\cite{2010-ICDM-Yang,2012-KDD-Borghol,2013-WSDM-Pinto}}) } & {\small\tt -}  \\ 
			 \cline{3-6}
			& &  {\tt TitleLen} & $\mathbb{N}$  & { \scriptsize Video title length in number of characters (\cite{li2016characterizing})} & {\small\tt n/l}  \\ 
			 \cline{3-6}
			& &  {\tt DescLen} & $\mathbb{Z}_+$  & { \scriptsize Video description length in number of characters (\cite{li2016characterizing})} & {\small\tt n/l}  \\ 
			 \cline{3-6}
			& &  {\tt UplDOW} & $(\mathbb{Z}_2)^7$   & { \scriptsize Day of the week of the video upload date ({\color{blue} new})} & {\small\tt -}  \\ 
			 \cline{3-6}
			& &  {\tt UplHour} & $[0,24)$  & { \scriptsize The hour of the video upload time ({\color{blue} new})} & {\small\tt -}  \\ 
			 \cline{2-6}
			
			 \cline{2-6}
			& \multirow{6}{*}{$\mathtt{API_{Sa}}$} & \multicolumn{4}{|l|}{\emph{  1.b. Static features from the video hosting service API about the video author}} \\
			 \cline{3-6}
			& &  {\tt AuthAge} & $\mathbb{Z}_+$  & { \scriptsize The author's age in number of days from her registration date (\cite{2012-KDD-Borghol,li2016characterizing}, \emph{\cite{2013-CWSM-Kupavskii})} } & {\small\tt n/l} \\ 
			 \cline{3-6}
			& &  {\tt AUplCnt} & $\mathbb{N}$  & { \scriptsize The number of videos uploaded by the author (\cite{2012-KDD-Borghol}, \emph{\cite{2011-WWW-Hong,2013-CWSM-Kupavskii}}) } & {\small\tt n/l} \\ 
			 \cline{3-6}
			& &  {\tt AViewSum} & $\mathbb{Z}_+$   & { \scriptsize The total time in seconds that viewers watched all the author's videos (\cite{2012-KDD-Borghol})} & {\small\tt n/l} \\ 
			 \cline{3-6}
			& &  {\tt FrndCnt} & $\mathbb{Z}_+$  & { \scriptsize The number of the author's friends  (\cite{2012-KDD-Borghol,li2016characterizing}, \emph{\cite{2013-CWSM-Kupavskii}})} & {\small\tt n/l} \\ 
			 \cline{3-6}
			& &  {\tt SubsCnt} & $\mathbb{Z}_+$  & { \scriptsize The number of the author's subscribers (\cite{2012-KDD-Borghol,li2016characterizing}, \emph{\cite{2013-CWSM-Kupavskii}})} & {\small\tt n/l} \\ 
			\hline
			\hline
			\multicolumn{2}{|l||} {\multirow{9}{*}{$\mathtt{API_D}$}} & \multicolumn{4}{|l|}{ \bf \emph{ 2. Dynamic features from the video hosting service API} } \\
			 \cline{3-6}
			\multicolumn{2}{|l||} {} & {\tt CommCnt} & $\mathbb{Z}_+$  & { \scriptsize The number of all comments on the video (\cite{2012-KDD-Borghol}, \cite{2013-WSDM-Figueiredo})}  & {\small\tt n/l,c/d}  \\ 
			 \cline{3-6}
			\multicolumn{2}{|l||} {} & {\tt LikeCnt} & $\mathbb{Z}_+$  & { \scriptsize The number of likes of the video (\cite{2012-KDD-Borghol})} & {\small\tt n/l,c/d}  \\ 
			 \cline{3-6}
			\multicolumn{2}{|l||} {} & {\tt DislCnt} & $\mathbb{Z}_+$  & { \scriptsize The number of dislikes of the video (\cite{2012-KDD-Borghol})} & {\small\tt n/l,c/d}  \\ 
			 \cline{3-6}
			\multicolumn{2}{|l||} {} & {\tt MinRat} & $\mathbb{Z}_5$  & { \scriptsize The minimum rating assigned to the video (\cite{2012-KDD-Borghol})} & {\small\tt -}  \\ 
			 \cline{3-6}
			\multicolumn{2}{|l||} {} & {\tt MaxRat} & $\mathbb{Z}_5$  & { \scriptsize The maximum rating assigned to the video (\cite{2012-KDD-Borghol})} & {\small\tt -}  \\ 
			 \cline{3-6}
			\multicolumn{2}{|l||} {} & {\tt AvgRat} & $[1,5]$  & { \scriptsize The average rating assigned to the video (\cite{2012-KDD-Borghol})} & {\small\tt -}  \\ 
			 \cline{3-6}
			\multicolumn{2}{|l||} {} & {\tt RatCnt} & $\mathbb{Z}_+$  & { \scriptsize The number of ratings assigned to the video (\cite{2012-KDD-Borghol})} & {\small\tt n/l,c/d}  \\ 
			 \cline{3-6}
			\multicolumn{2}{|l||} {} & {\tt Update} & $\mathbb{Z}_+$  & { \scriptsize The number of days passed from the last update date (\cite{2012-KDD-Borghol})} & {\small\tt n/l}  \\ 
			\hline
			
		\end{tabular}
	}
\end{table*}

\begin{table*}
	\centering
	\caption{List of features from {\tt LOG} and {\tt WEB} that are used to predict video popularity ({\color{blue}all of them are  novel in the context of video popularity prediction}).
		\label{FactorTable_LOGWEB}}{%	
		\begin{tabular}{|l|l||p{1.5cm}|c|p{6.5cm}|p{1.1cm}|} 
			\hline
			\multicolumn{2}{|c||} {\bf Feature set }  & \centering  {\bf Name } & {\bf Values } & \multirow{2}{*}{\bf Description (all of them are {\color{blue}new})} & \multirow{2}{*}{\bf Modes}\\ 
			\multicolumn{2}{|c||} {\bf $\mathbf{\Psi}$}  & \centering  {\bf $\psi$} & {\bf $\mathbb{D}_\psi$ } &  &    \\ 
			\hline
			\hline
			\multirow{7}{*}{{\tt LOG}} &   & \multicolumn{4}{|l|}{ \bf \emph{ 3. Dynamic features from logs of Yandex} } \\
			 \cline{2-6}
			
			 \cline{2-6}
			& \multirow{4}{*}{$\mathtt{LOG_S}$}   & \multicolumn{4}{|l|}{\emph{  3.a. Dynamic features from search logs of Yandex}}   \\
			 \cline{3-6}
			& &  {\tt ShowURL} & $\mathbb{Z}_+$  & { \scriptsize The number of shows of the video URLs on SERP } & {\small\tt n/l,c/d} \\ 
			 \cline{3-6}
			& &  {\tt ClickURL} & $\mathbb{Z}_+$  & { \scriptsize The number of clicks on the video URLs on SERP } & {\small\tt n/l,c/d} \\ 
			 \cline{3-6}
			& &  {\tt CTR} & $[0,1]$  & { \scriptsize The click--through rate of the video URLs on SERP } & {\small\tt c/d} \\ 
			 \cline{3-6}
			
			 \cline{2-6}
			& \multirow{2}{*}{$\mathtt{LOG_B}$}   & \multicolumn{4}{|l|}{\emph{  3.b. Dynamic features from browsing logs of Yandex}} \\
			 \cline{3-6}
			& &  {\tt BrowVisit} & $\mathbb{Z}_+$  & { \scriptsize The number of visits of the video URLs registered in browsing logs } & {\small\tt n/l,c/d} \\ 
			\hline
			\hline
			\multirow{21}{*}{{\tt WEB}} &   & \multicolumn{4}{|l|}{ \bf \emph{ 4. Dynamic features from the Web } } \\
			 \cline{2-6}
			
			 \cline{2-6}
			& \multirow{17}{*}{$\mathtt{WEB_{ag}}$}   & \multicolumn{4}{|l|}{\emph{  4.a. Dynamic aggregated features from the Web}} \\
			 \cline{3-6}
			& &  {\tt EmbCnt} & $\mathbb{Z}_+$  & { \scriptsize The number of all embeds of the video } & {\small\tt n/l,c/d} \\ 
			 \cline{3-6}
			& &  {\tt EmbHCnt} & $\mathbb{Z}_+$  & { \scriptsize The number of all hosts with embeds of the video } & {\small\tt n/l,c/d} \\ 
			 \cline{3-6}
			& &  {\tt MaxEPerH} & $\mathbb{Z}_+$  & { \scriptsize The maximum number of embeds of the video per host } & {\small\tt n/l,c/d} \\ 
			 \cline{3-6}
			& &  {\tt AvgEPerH} & $\mathbb{R}_+$  & { \scriptsize The average number of embeds of the video per host } & {\small\tt n/l,c/d} \\ 
			 \cline{3-6}
			& &  {\tt MaxEPerP} & $\mathbb{Z}_+$  & { \scriptsize The maximum number of embeds of videos per page } & {\small\tt n/l,c/d} \\ 
			 \cline{3-6}
			& &  {\tt AvgEPerP} & $\mathbb{R}_+$  & { \scriptsize The average number of embeds of videos per page } & {\small\tt n/l,c/d} \\ 
			 \cline{3-6}
			& &  {\tt FirstEmb} & $\mathbb{Z}_+$  & { \scriptsize The number of days passed since the first embed of the video} & {\small\tt n/l} \\ 
			 \cline{3-6}
			& &  {\tt LastEmb} & $\mathbb{Z}_+$  & { \scriptsize The number of days passed since the last embed of the video} & {\small\tt n/l} \\ 
			 \cline{3-6}
			& &  {\tt AvgEmb} & $\mathbb{R}_+$  & { \scriptsize The average number of days passed since any embed of the video} & {\small\tt n/l} \\ 
			 \cline{3-6}
			& &  {\tt LinkCnt} & $\mathbb{Z}_+$  & { \scriptsize The number of all links to the video } & {\small\tt n/l,c/d} \\ 
			 \cline{3-6}
			& &  {\tt LinkHCnt} & $\mathbb{Z}_+$  & { \scriptsize The number of all hosts with links to the video } & {\small\tt n/l,c/d} \\ 
			 \cline{3-6}
			& &  {\tt MaxLPerH} & $\mathbb{Z}_+$  & { \scriptsize The maximum number of links to the video per host } & {\small\tt n/l,c/d} \\ 
			 \cline{3-6}
			& &  {\tt AvgLPerH} & $\mathbb{R}_+$  & { \scriptsize The average number of links to the video per host } & {\small\tt n/l,c/d} \\ 
			 \cline{3-6}
			& &  {\tt FirstLink} & $\mathbb{Z}_+$  & { \scriptsize The number of days passed since the day of the first link } & {\small\tt n/l} \\ 
			 \cline{3-6}
			& &  {\tt LastLink} & $\mathbb{Z}_+$  & { \scriptsize The number of days passed since the video was linked last time} & {\small\tt n/l} \\ 
			 \cline{3-6}
			& &  {\tt AvgLink} & $\mathbb{R}_+$  & { \scriptsize The average number of days passed since there was any link to the video } & {\small\tt n/l} \\ 
			 \cline{2-6}
			 \cline{2-6}
			& \multirow{3}{*}{$\mathtt{WEB_{nag}}$}   & \multicolumn{4}{|l|}{\emph{  4.b. Dynamic non-aggregated features from the Web}}  \\
			 \cline{3-6}
			& &  {\tt EmbedHost} & $\mathbb{E}$  & { \scriptsize The host list with embed timestamps of the video (preprocessed into the  outcomes of the LIM) } & {\small\tt n/l,c/d} \\ 
			 \cline{3-6}
			& &  {\tt LinkHost} & $\mathbb{L}$  & { \scriptsize The host list with link timestamps of the video (preprocessed into the  outcomes of the LIM) } & {\small\tt n/l,c/d} \\ 
			\hline
		\end{tabular}
	}
\end{table*}

We summarized and described all features used in our work in Tables~\ref{FactorTable_API} and~\ref{FactorTable_LOGWEB}.
The sets $\mathbf{\Psi}$ that the feature belongs to are specified in the column ``Feature set". The feature value space $\mathbb{D}_\psi$ is described in the column ``Values"\footnote{We remind the standard number set notations: 
	$\mathbb{N}$ is the set of natural numbers, i.e. $\{1,2,..\}$; 
	$\mathbb{Z}_+$ is the set of non-negative integer numbers, i.e. $\{0,1,2,..\}$; 
	$\mathbb{R}_+$, is the set of non-negative real numbers, i.e. $[0,+\infty)$, 
	$\mathbb{Z}_p, p\in\mathbb{N},$ is the finite set of numbers $\{0,1,..,p-1\}$.
}.
For each integer or real valued feature that has $\gg 1$ values, we consider both its actual value and its \emph{logarithmic} transformation (in terms of $\log(x)\stackrel{def}{=}\log_2(x+1)$). The logarithmic transformation of the feature $\varphi\in\mathbf{\Phi}, \: \mathbb{D}_\varphi\subset\mathbb{R},$ we denote by {\tt log($\varphi$)}. For instance, {\tt TitleLen} is the actual number of characters in the title of a video, and {\tt log(TitleLen)} is the logarithm of that number. The presence  of the mark  ``{\tt n/l}" in Tables~\ref{FactorTable_API}\&\ref{FactorTable_LOGWEB}, column ``Modes", means that the marked feature is used in popularity prediction both in transformed and non-transformed forms.
The column ``Modes" of Tables~\ref{FactorTable_API}\&\ref{FactorTable_LOGWEB} contains also marks ``{\tt c/d}". The presence of the mark means that the corresponding feature possesses the integral property. 
It means that the feature value at a time moment $t\in\mathbb{T}$ is equal to the sum of elementary features
over some time period. In our work, we use two options: the value calculated during the entire period $[0, t)$ (\emph{cumulative}) and the value calculated during the last day $[t -1, t) (\emph{daily})$\footnote{It is possible to derive many other combinations like values calculated during the two previous days $[t -2, t)$ and so on. It is not the goal of our research, so we took just 2 different variants. A deeper investigation of other combinations could be regarded as future work.}. In order to distinguish cumulative and daily feature values, we add to the feature name  suffixes ``{\tt[c]}" and ``{\tt[d]}", correspondingly. For instance, {\tt EmbCnt[c]} is the number of embeds that a video received during its existence up to the current time moment $t_c$, whereas {\tt EmbCnt[d]} is the number of embeds that the video received during the last day before the current time moment $t_c$, that is $[t_c -1, t_c)$.

We split the features from the hosting provider into the dynamic feature set $\mathtt{API_D}$ and  the static feature set $\mathtt{API_S}$. The static feature set consists of the set $\mathtt{API_{Sv}}$ of features about the video and the set $\mathtt{API_{Sa}}$ of features about the author of the video. The features from the sets {\tt LOG} and {\tt WEB} are dynamic. The feature set  {\tt LOG} extracted from the logs of Yandex, we split into the features from the search logs $\mathtt{LOG_S}$ and the features from the browsing logs $\mathtt{LOG_B}$. The feature set {\tt WEB} extracted from the publicly available web resources are based on monitoring both embeds of and links to the videos. We split them into aggregated features $\mathtt{WEB_{ag}}$ and non-aggregated features $\mathtt{WEB_{nag}}$.

Non-aggregated features are introduced only to use them within the linear influence model described in Section~\ref{sec_Models}. The difference between aggregated and non-aggregated features consists in the following. An aggregated feature is a feature that aggregates the information about a number of elementary features, that are called \emph{non-aggregated features}. For instance, the fact of the video's embed(s) at a specific web site (host) can be represented by an elementary non-aggregated feature. Usually,  because of their large number, such features are aggregated in a small number of features with each of them representing some aspect (e.g., the total number of hosts that have at least one embed of or a link to the video). In our paper, for a non-aggregated embed feature, we have feature value space $\mathbb{E}$ that has the form $
\mathbb{E} = (2 ^ {\mathbb{T} \times \mathbb{N}})^{c_e},$ where $c_e = |\mathcal{H}'|$ is the number of used hosts and each element $(t,n)\in\mathbb{T} \times \mathbb{N}$ corresponds to the event that a particular host embeds a video $n$ times at its $n$ pages at the $t$-th day since the video creation. The feature value space $\mathbb{L}$ in the case of links is of the same form as $\mathbb{E}$.

Finally, the table provides the information about which features were previously investigated in the literature devoted to video popularity (column ``Description  (previously used  or new)"\footnote{The \emph{italic} style of the reference means that the feature was just investigated for some other task,  but was not used to predict popularity. }). From that one can learn that feature sets $\mathtt{API_{Sa}}$ and $\mathtt{API_D}$ are entirely previously investigated, while the feature set $\mathtt{API_{Sv}}$ contains features that are not previously investigated. We unite previously investigated features of the set $\mathtt{API_{Sv}}$ in the set denoted by $\mathtt{API_{Svb}} $ {\tt = \{Dur, Cat, TitleLen, DescLen\}}. Then all previously investigated features are united in the set denoted by $\mathtt{BASE.lit} = \mathtt{API_{Svb}}\cup\mathtt{API_{Sa}}\cup\mathtt{API_D}$. So, the set {\tt BASE.lit} is our first baseline feature set  (\emph{related work based baseline}). At the same time, we will consider all features that we could extract from the data acquired from the hosting provider's API (as of January 2014) as our second baseline feature set (\emph{API baseline}), that is the set $\mathtt{API} = \mathtt{API_S}\cup\mathtt{API_D}$.
At last, we define a short synonymous notation for the set of all features $\mathtt{ALL} = \mathtt{API}\cup\mathtt{WEB}\cup\mathtt{LOG}$.

\section{Models}
\label{sec_Models}
\subsection{General model}
Since our main goal is to compare different features, we use the same prediction model for all features. We considered  a state-of-the-art  \emph{Friedman's gradient boosting decision tree model} \cite{2001-AS-Friedman} and a traditional \emph{linear regression model}. The model's characteristics are described more precisely at the beginning of Section~\ref{sec_ExpsSetup}. For non-aggregated features we implement the {\bf linear influence model}, which is described further in this subsection. The predictions of the linear influence model are combined with other aggregated features, that is, we use them as additional features of decision tree models.

\subsection{Linear influence model}
The \emph{linear influence model} (\emph{LIM}) was introduced in \cite{2010-ICDM-Yang}, where it was used to predict diffusion of hashtags over Twitter network and diffusion of memes (short textual phrases) over news articles and blog posts. The model's implementation aspects and modifications were discussed later in  \cite{2013-AAAI-Wang}. The main advantage of the model consists in that it does not require any knowledge of the network structure where the information is spreading. 

In our work, we consider a video $v \in \mathcal{V}$ as some infection. It diffuses through implicit network of users and web sites (or \emph{hosts}). A video infects users when they watch it and it infects hosts if they contain web pages with an embed of the video or a link to the video web page. So, the diffusion network could be represented as a bipartite graph, where the first layer of nodes is a set of users $\mathcal{U}$ and the second layer of nodes is a set of web sites (hosts) $\mathcal{H}$.

Then, the linear influence model states that each node $h\in\mathcal{H}$ possesses a particular influence function $I_h: \{0,..,(L-1)\} \rightarrow \mathbb{R}_+$, where $L$ is the size of the function \emph{domain}. The value $I_h(t)$ is equal to the number of nodes from $\mathcal{U}$ that will be infected  by the node $h$ during the $t$-th day after the node $h$ was infected, that is, during the time period $[t_h + t - 1, t_h + t )$, where $t_h\in \overline{\mathbb{T}}$ is the time moment when the node $h$ was infected. Then the number of views of a video in a particular day equals to the outcomes of a sum of the  influence functions of previously infected hosts (see \cite{2010-ICDM-Yang} for details).

Thus, on the one hand, the number of views of a video $v \in \mathcal{V}$ is exactly the number of times when the nodes from the set $\mathcal{U}$ were infected. On the other hand, the operating company cannot observe particular user infections, but can observe a subset of web sites $\mathcal{H}' \subset \mathcal{H}$ and fix the time when some of them embed the video (or create a link to the video web page), and hence become infected. 

To the best of our knowledge, we are the first who introduced the application of the linear influence model to a bipartite graph, where each part has  its own criteria for infection and where a prediction of infection spread of the one part is made based on the infection spread observed in the other.
Since we investigated the case when negative values of influence functions are allowed (sometimes, the fact of an embed at a particular highly specialized host may indicate that the video will not be popular), hence the LIM prediction problem is reduced to the least squares problem with a sparse matrix. 
Thus, the LIM allows us to use video features in non-aggregated forms. The outcomes predicted by LIM are used as features of our general model (see Section~\ref{sec_ExpsSetup}).

\section{Experimental setup}
\label{sec_ExpsSetup}
\subsection{Model settings}
As it was stated above, the main objective of our study is an investigation of the wide range of available data for the task of video popularity prediction. Thus, we use the same machine learning algorithm for all experiments conducted in our work, except for the case of non-aggregated  features $\mathtt{WEB_{nag}}$, where we use the LIM (see Section~\ref{sec_Models}). In all described experiments of our work, we used a proprietary implementation of the gradient boosted decision tree-based machine learning algorithm \cite{2001-AS-Friedman} with 1000 iterations and 1000 trees, which appeared to be the best settings on the validation data.
During our experimentation, we also utilized the traditional \emph{linear regression model}, however, this model demonstrated a considerably worse performance with respect to the decision trees (by a minimum margin of $10\%$ in our experiments).

%{\bf Linear influence model.}
As it was stated in Section~\ref{sec_Models}, the linear influence model has the following parameters: the size $L$ of the influence function domain, the size $|\mathbb{T}|$ of the learning time period, the learning set $\mathcal{H}'$ of hosts, the training set $\mathcal{V}'$ of videos (infections).
Since prediction of video popularity with LIM model is computationally expensive (though the problem matrix is sparse, what seriously decreases the number of elementary operations), we implemented the LIM solver on a distributed cluster system with the proprietary  MapReduce technology \cite{2008-CACM-Dean} that allowed to train the model using more than $8\cdot 10^6$ training videos  in acceptable time.

We conducted a series of experiments in order to determine how the LIM parameters affect the prediction quality. We found that for different zones of the set of target days $\mathbb{T}$ the model has different optimal parameters $L$ and $|\mathbb{T}|$.  Thus, we learn $12$ LIMs on the Dataset\#2 with different values of the parameters both for embed data and for link data. As a result, we obtained 6 influence functions  per host (with the top-1280 hosts ranked by total number of embeds ($|\mathcal{H}'|=1280$), the domain size $L \in \{ 1, 10, 20 \}$, and the sizes $|\mathbb{T}|$ were chosen equal to optimal values per each $L$). The outcome of each of these functions is a quadruple (non-/log-transformed cumulative/daily popularity). Thus, 12 LIM functions contributed $48$ features studied in the experiments described further in Section~\ref{sec_ExpsRes}.

\begin{table}
	\centering
	\caption{Baseline comparison in terms of the average normalized RMSE  over first 14 days since video creation (in \% with respect to {\tt API}).\label{BaselinesTable}}{%
		
		\begin{tabular}{|l||c|c||c|c|} 
			\hline
			\multirow{3}{*}{Features} & \multicolumn{4}{|c|}{Targets ($\theta\in\mathbf{\Theta}$)} \\ 
			\cline{2-5}
			
			& \multicolumn{2}{|c||}{{\tt log}-transformed}  & \multicolumn{2}{|c|}{non-transformed} \\ 
			\cline{2-5}
			& cumulative & daily & cumulative & daily \\ 
			\hline
			\hline
			{  \tt API}	  & $\mathbf{0.356}  $ & $\mathbf{0.435}	 $ & $\mathbf{0.822}	 $ & $\mathbf{0.91} $ \\
			\hline
			{  \tt BASE.lit}	  & $\mathbf{+0.79\%}	 $ & $+3.02\%	 $ & $\mathbf{+0.81\%}	 $ & $+3.92\% $ \\
			\hline
			\hline
			{  $\mathtt{API_{Sv}}$ }	  & $+154.96\%	 $ & $+112.91\%	 $ & $+21.68\%	 $ & $+9.85\% $ \\
			\hline
			{  $\mathtt{API_{Sa}}$ }	  & $+38.44\%	 $ & $+33.08\%	 $ & $+11.82\%	 $ & $+6.78\% $ \\
			\hline
			{  $\mathtt{API_D}$ }	 & $+53.54\%	 $ & $+26.99\%	 $ & $+2.26\%	 $ & $+1.73\% $ \\
			\hline
			\hline
			{  $\mathtt{API_{Sa}}\cup \mathtt{API_D}$ }	  & $+4.03\%	 $ & $+2.95\%	 $ & $\mathbf{-0.3\%}	 $ & $\mathbf{-0.63\%} $ \\
			\hline
			{  $\mathtt{API_{Sv}}\cup \mathtt{API_D}$ }	  & $+33.75\%	 $ & $+16.45\%	 $ & $+2.67\%	 $ & $+1.62\% $ \\
			\hline
			{  $\mathtt{API_{Sv}}\cup \mathtt{API_{Sa}}$ }  & $+33.93\%	 $ & $+29.53\%	 $ & $+12.62\%	 $ & $+6.64\% $ \\
			\hline
		\end{tabular}
	}
\end{table}

\subsection{Target and target days}
In accordance with our problem statement (Section~\ref{sec_ProblemStat}), we run our models for each of the target days $t_t\in\mathbb{T}^*$, i.e., for $1,\ldots, 14$ days since video creation time, and for all targets from $\mathbf{\Theta}$, see Eq.~(\ref{eq:Theta_def}). We address both types of prediction tasks:
\begin{itemize} 
	\item the current popularity prediction, i.e., where the current time moment equals to the target one: $t_c=t_t$;
	\item the future popularity prediction with forecast days $1, \ldots, 13$, i.e., where the current time moment is lower than the target one by an increment $\delta$: $t_c = t_t - \delta$, $\delta=1,2,\ldots,t_t-1$.  
\end{itemize}

\subsection{Performance measures}
\label{subsec_PerfMeas}
Since, the values of RMSE are not normalized, they vary considerably depending on the target time $t_t$\footnote{It is caused by a large difference in the number of views for different days. The variation of non-normalized RMSE could be seen in Figure~\ref{img_AllExp} at the next experiment discussions.} and that could make it difficult to interpret results. Therefore, in the major part of the results we will normalize the RMSE values by the RMSE values of the baseline $\mathtt{BASE.avg}$ (see Section~\ref{subsec_ExpRes_base} for its description) for each set of features $\mathbf{\Psi}$ and for each $t_t \in \mathbb{T}^*$ and $t_c = t_t - \delta$, $\delta=1,2,\ldots,t_t-1$:
$$
	\mathrm{nRMSE}(\mathbf{\Psi}, t_c; \theta, t_t) = \frac{\mathrm{RMSE}(\mathbf{\Psi}, t_c; \theta, t_t)}{\mathrm{RMSE}(\mathtt{BASE.avg}, t_c; \theta, t_t)}.
$$
After normalization one could obtain the average normalized RMSE (AnRMSE) over all target days $\mathbb{T}^*$, e.g., for the current popularity prediction case $t_c = t_t$:
$$
	\mathrm{AnRMSE}(\mathbf{\Psi}; \theta) = |\mathbb{T}^*|^{-1}\cdot\sum\limits_{t_t\in\mathbb{T}^*} \mathrm{nRMSE}(\mathbf{\Psi}, t_t; \theta, t_t).
$$

As the second performance measure, we will use \emph{normalized discounted cumulative gain} \cite{2002-TOIS-Jarvelin} (\emph{NDCG@100}) over the top-100 results predicted as the most popular by our methods. We consider this measure as the most relevant to our study as it directly reflects the profit that an operating company (and, especially, a search engine) may receive from a high quality popularity prediction mechanism. In our work, the gain of a ranked video equals to $1/(\mathrm{pos}+1)$ (where $\mathrm{pos}$ is the position of the video in the  (ideal) list, in which videos are ranked by the actual target value) by the target value for the first $100$ most popular videos, and equals to $0$, if the ranked video does not belong the the top-100 videos of that ideal ranking.

\subsection{Test and training data sets split} 
We split the Dataset\#1 randomly into three equal parts as it is done in \cite{2010-CACM-Szabo,2013-WSDM-Pinto,2013-WSDM-Ahmed} for video popularity prediction. The first part is used as the test data, 
the second one serves as the training data and the third part is used as the validation set.
We repeated this procedure 20 times in order to apply the
\emph{paired two-sample t-test} and measure the significance level of the
obtained results. All  differences in the presented results have  p-value $<0.05$.
{
\sloppy
	
}
\section{Experiment Results}
\label{sec_ExpsRes}

Our results for forecasting (i.e., $t_c<t_t$) are very similar to the results for current popularity prediction (i.e., $t_c=t_t$). Hence, in the subsections where we compare feature sets and models (i.e., in Sections~\ref{subsec_ExpRes_base}, \ref{subsec_ExpRes_WebLog_vs_API}, \ref{subsec_ExpRes_APIparts}, and~\ref{subsec_ExpRes_ranking}), we describe only the results for the latter one, which we consider a novel and more relevant task for our study. On the contrary, the analysis of the performance of future popularity prediction with different forecasting horizons and its comparison with the one of current popularity prediction are done in Section~\ref{subsec_ExpRes_delay}.

\begin{table}
	\centering
	
	\caption{Comparison of API, Web and Log feature sets in terms of the average normalized RMSE  over first 14 days since video creation (in \% with respect to {\tt API}).\label{AllExpTable}}{%
		
		\begin{tabular}{|l||c|c||c|c|} 
			\hline
			\multirow{3}{*}{Features} & \multicolumn{4}{|c|}{Targets ($\theta\in\mathbf{\Theta}$)} \\ 
			\cline{2-5}
			
			& \multicolumn{2}{|c||}{{\tt log}-transformed}  & \multicolumn{2}{|c|}{non-transformed} \\ 
			\cline{2-5}
			& cumulative & daily & cumulative & daily \\ 
			\hline
			\hline
			{  \tt API	}  & $ 0.356 (0\%)	 $ & $ 0.435 (0\%)	 $ & $ 0.822 (0\%)	 $ & $ 0.91 (0\%) $ \\
			\hline
			{   $\mathtt{API}\cup \mathtt{LOG}$	} & $ -1.4\%	 $ & $ -1.53\%	 $ & $ -2.61\%	 $ & $ -0.27\% $ \\
			\hline
			{   $\mathtt{API}\cup \mathtt{WEB}$ } & $ -1.19\%	 $ & $ -0.8\%	 $ & $ -10.15\%	 $ & $ -4.36\% $ \\
			\hline
			{  \tt ALL } & $ \mathbf{-2.37\%}	 $ & $  \mathbf{-2.11\%}	 $ & $  \mathbf{-10.7\%}	 $ & $  -4.28\% $ \\
			\hline
			{   $\mathtt{ALL}\setminus \mathtt{WEB_{nag}}$	} & $  \mathbf{-2.35\%}	 $ & $  \mathbf{-2.09\%}	 $ & $  \mathbf{-10.72\%}	 $ & $  -4.75\% $ \\
			\hline
			{   $\mathtt{API}\cup \mathtt{WEB_{ag}}$	} & $ -1.17\%	 $ & $ -0.8\%	 $ & $ -10.11\%	 $ & $ \mathbf{-5.29\%} $ \\
			\hline
			\hline
			{  \tt LOG	 } & $ +156\%	 $ & $ +106.94\%	 $ & $ +18.29\%	 $ & $ +8.49\% $ \\
			\hline
			{  \tt WEB	} & $ +142.21\%	 $ & $ +98.12\%	 $ & $ +8.59\%	 $ & $ +3.74\% $ \\
			\hline
			{   $\mathtt{WEB}\cup \mathtt{LOG}$	 } & $ +132.74\%	 $ & $ +89.07\%	 $ & $ +4.64\%	 $ & $ +3.75\% $ \\
			\hline
			{   $\mathtt{WEB_{ag}}\cup \mathtt{LOG}$	} & $ +132.74\%	 $ & $ +89.07\%	 $ & $ +4.46\%	 $ & $ +3.92\% $ \\
			\hline
			{  $\mathtt{WEB_{nag}}$	 } & $ +143.01\%	 $ & $ +99.72\%	 $ & $ +9.18\%	 $ & $ +6.05\% $ \\
			\hline
			{  $\mathtt{WEB_{ag}}$	 } & $ +142.21\%	 $ & $ +98.12\%	 $ & $ +9.13\%	 $ & $ +4.89\% $ \\
			\hline
			
		\end{tabular}
	}
\end{table}

\subsection{Baselines} 
\label{subsec_ExpRes_base}
Before the start of our investigation of  the web and internal log data utility, we describe our baseline methods.
We have implemented the following baselines:
\begin{itemize}
 
	\item the predictions based on all API features ($\mathtt{API}$);
 
	\item the predictions based on the API features used previously in the related studies ($\mathtt{BASE.lit}$) (see Sections~\ref{sec_RelWork} and \ref{sec_Features});
 
	\item the naive average prediction model,  which, for each video from the test data set, predicts the target value as the average of the corresponding target values on the training data set for all videos.  This baseline is denoted by $\mathtt{BASE.avg}$. 
 
\end{itemize}

The average normalized RMSE values for the baseline methods are presented in Table~\ref{BaselinesTable} (except for $\mathtt{BASE.avg}$ whose AnRMSE is equal to $1$). In the table, we compare the strength of feature sets $\mathtt{BASE.lit}$, $\mathtt{API_{Sv}}$, $\mathtt{API_{Sa}}$, $\mathtt{API_D}$, $\mathtt{API_{Sa}}\cup \mathtt{API_D}$, $\mathtt{API_{Sv}}\cup \mathtt{API_D}$, and $\mathtt{API_{Sv}}\cup \mathtt{API_{Sa}}$ by looking at the relative change of the metric against the full API feature set $\mathtt{API}$.
On the one hand, one could see that the set $\mathtt{API}$ outperforms the set $\mathtt{BASE.lit}$  for all targets and mainly in daily views, both transformed and not. On the other hand, the set $\mathtt{API_{Sa}}\cup \mathtt{API_D}$  makes the major contribution to the quality of prediction of the API set: it has no noticeable difference  for non-transformed targets and loses a little bit  on the {\tt log}-targets. Thus, further in the paper we will use only the best baseline using all API features.

\subsection{Web and logs vs API}
\label{subsec_ExpRes_WebLog_vs_API}
The feature set {\tt All} outperforms the API features. Thus, the web and log data notably improve the video popularity prediction quality in terms of all targets. This result could be seen in Table~\ref{AllExpTable} and serves as the answer to the {\bf RQ1}. In the same table, we compare all other feature  sets with the API  baseline and with each other.

\begin{figure}
	\centering
	\includegraphics[width=\textwidth]{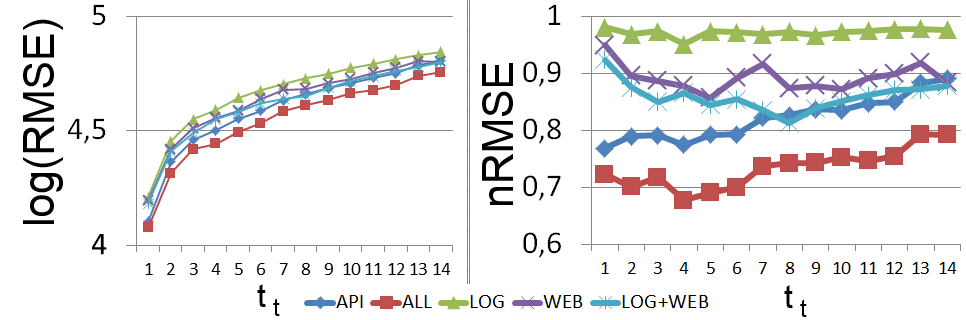}
	
	\caption{ Comparison of API, Web and Log feature sets in terms of the RMSE (on the left) and the normalized RMSE (on the right) for each of the first 14 days since video creation for the target $\mathtt{Views[c]}$.}
	
	\label{img_AllExp}
\end{figure}

One could see that the absence of API data has the most dramatic effect for $log$-transformed targets and only slightly reduces the video popularity prediction quality for non-transformed targets ({\tt API} vs $\mathtt{WEB}\cup \mathtt{LOG}$). We conclude that the Web and the Log data could not completely replace API data for the purposes of the video popularity prediction in terms of exact values aggregated for 14 days. A completely different picture is observed for specific days (see further) and for ranking measures (see Section~\ref{subsec_ExpRes_ranking}). 
At the same time one can see that the embed and link data ({\tt WEB})  improve the quality better than the internal log data ({\tt LOG}) of the operating company both independently ({\tt WEB} vs {\tt LOG}) and together with other features ($\mathtt{API}$ vs $\mathtt{API}\cup \mathtt{WEB}$ vs $\mathtt{API}\cup \mathtt{LOG}$), ($\mathtt{ALL}$ vs $\mathtt{API}\cup \mathtt{WEB}$ vs $\mathtt{API}\cup \mathtt{LOG}$).
One could also see that the non-aggregated web features ($\mathtt{WEB_{nag}}$) used noticeably improve prediction quality of the aggregated ($\mathtt{WEB_{ag}}$) for non-transformed targets: by 1.14\% for daily and by 0.54\% for cumulative views ($\mathtt{WEB}$ vs $\mathtt{WEB_{ag}}$). It is an important result given that we study the potential of an operating company to work in the absence of access to APIs.

In Figure~\ref{img_AllExp}, we demonstrate the values of the RMSE and the normalized RMSE measures per each day for the non-transformed cumulative target for the feature sets {\tt API}, {\tt ALL}, {\tt LOG}, {\tt WEB}, and $\mathtt{WEB}\cup \mathtt{LOG}$. 
Wherein, on the one hand, one could see that the API features lose quality with the growth of the number of the days passed since  the video upload. The same situation is observed with the whole feature set $\mathtt{ALL}$ (which includes API features). On the other hand, the Web and Log data improve the quality of prediction of exact values of views with the growth of $t_t$ at least to the end of the first week of the video existence, and \emph{are able to compensate for the absence of API starting from the 8th day}. It is the partial answer to the {\bf RQ2} (Section~\ref{subsec_ExpRes_ranking} has an extended and more definitive answer to {\bf RQ2}).

\subsection{Replacing parts of API with Web/Log data}
\label{subsec_ExpRes_APIparts}
We split the API feature set into the following groups  (see the notations in Table~\ref{FactorTable_API}): 
\begin{itemize}
	\item temporal context, {\tt tc  = \{UplHour, Update, UplDOW\}}; 
	\item static video properties, {\tt sv= \{Cat, Dur, TitleLen, DescLen\}};
	\item user feedback, {\tt uf = \{Min/Max/AvgRat, Like/DislCnt, RatCnt, CommCnt\}};
	\item author rating, { \tt ar = \{AuthAge, AUplCnt, AViewSum\}};
	\item social environment, {\tt se = \{FrndCnt, SubsCnt\}}. 
\end{itemize}
We measure the quality for the feature sets $(\mathtt{API} \setminus \mathtt{group})$ and $(\mathtt{ALL} \setminus \mathtt{group})$ by removing each $\mathtt{group}\in \{\mathtt{tc}, \mathtt{sv}, \mathtt{uf}, \mathtt{ar}, \mathtt{se}\}$ from {\tt API} and {\tt ALL} feature sets respectively.

\begin{table}
	\centering
 
	\caption{Ablation of feature groups from API set with their further replacement by the Web and Log features.  All results are in terms of the average normalized RMSE  over first 14 days since video creation (in \% with respect to the feature set {\tt API}).\label{GroupsExpTable}}{%
		
	\begin{tabular}{|l||c|c||c|c|} 
		\hline
		\multirow{3}{*}{$\mathtt{group}$} & \multicolumn{4}{|c|}{ Cumulative number of views ($\theta\in\mathbf{\Theta}$)} \\ 
		\cline{2-5}
		
		& \multicolumn{2}{|c||}{{\tt log}-transformed}  & \multicolumn{2}{|c|}{non-transformed} \\ 
		\cline{2-5}
		& {  $\mathtt{API} \setminus \mathtt{group} $} & 	 {  $\mathtt{ALL} \setminus \mathtt{group} $} 
		& {  $\mathtt{API} \setminus \mathtt{group} $} & 	 {  $\mathtt{ALL} \setminus \mathtt{group} $} \\
		
		\hline
		\hline
		{ \tt tc}& 	$+0.43\% $ & 	$\mathbf{-1.97}\% $ & 	$+0.02\% $ & 	$\mathbf{-10.8}\% $ \\
		\hline
		{ \tt sv}& 	$+3.95\% $ & 	$\mathbf{+1.28}\% $ & 	$-0.8\% $ & 	$\mathbf{-10.81}\% $ \\ 
		\hline
		{ \tt uf}& 	$+27.6\% $ & 	$+20.27\% $ & 	$+12.11\% $ & 	$\mathbf{-2.87}\% $ \\ 
		\hline
		{ \tt ar}& $+18.6\% $ & 	$+14.59\% $  &	$+1.14\% $ & 	$\mathbf{-9.91}\% $ \\
		\hline
		{ \tt se}& 	$+3.73\% $ & 	$\mathbf{+0.82}\% $ & 	$+1.56\% $ & 	$\mathbf{-9.53}\% $ \\ 
		\hline
		
	\end{tabular}
}
\end{table}

The obtained results for cumulative views\footnote{the results for daily views are similar} are presented in Table~\ref{GroupsExpTable}, where the columns ``$\mathtt{API} \setminus \mathtt{group} $" show how much the prediction quality falls when a particular group becomes unavailable via API, and the columns ``$\mathtt{ALL} \setminus \mathtt{group} $" show how much the prediction quality restores when we replace this group by the Web and the Log features. It is seen from the table that the absence of the \emph{user feedback} feature group has the most dramatic effect for both targets. The absence of the \emph{temporal context}, \emph{static video properties}, and \emph{social environment} feature groups has no significant consequences and could be easily replaced by the Web and the Log feature sets without any loss in the prediction quality. Finally, we conclude that \emph{the Web and Log data could compensate for the absence of any of the studied parts of reliable API data without any loss in the quality} and even with a solid profit, in fact. It is the answer to the {\bf RQ3}.

\subsection{Delay in data crawling}
\label{subsec_ExpRes_delay}
\begin{figure}
	\centering
	\includegraphics[width=\textwidth]{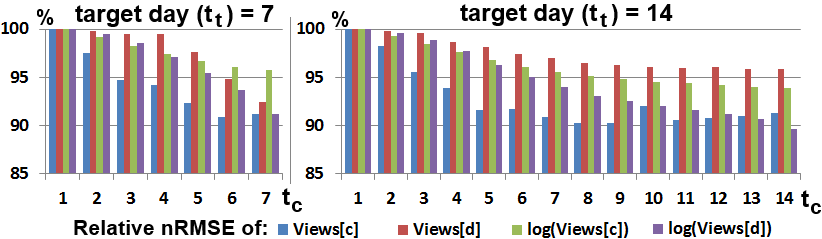}
 
	\caption{Prediction quality change of the feature set $\mathtt{ALL}$  with respect to different values of the current day $t_c$: from the first day since video creation to the target day $t_t$ (7 or 14)  (in \%  with respect to nRMSE of the data observed in the first day).}
 
	\label{img_ShiftExp}
\end{figure}

The prediction quality is also affected by the delay in data crawling. In order to study this influence, we fix the target day $t_t$ and measure the quality for the feature set $\mathtt{ALL}$, considering each day from the first day since video creation to this target day $t_t$ as the current day $t_c$, i.e., $t_c  = t_t - \delta, \delta  =0,..,13$. In other words, we predict the current video popularity at the day $t_t$, as if we are in the past with data collected at the day $t_c$. The relative values of nRMSE (in terms of \% with respect to the one for $t_c=1$) are presented in Fig.~\ref{img_ShiftExp} for $t_t = 7$ and~$14$, for each target $\theta\in\Theta$. 

We see that the shorter the crawling delay $\delta$ the better the prediction performance. 
However,  for the target day $t_t=7$, even a delay in one day leads to   a noticeable quality drop (up to $2.5\%$ for daily views), while, for the target day $t_t=14$, such delay leads to a less dramatic  quality drop (up to $1\%$ for daily views). We conclude that \emph{the speed of crawling (both of API, and of WEB data) has a strong influence on the popularity prediction performance, and the closer the prediction target day to the video creation moment, the more critical this influence is}.

\subsection{Ranking metric quality}
\label{subsec_ExpRes_ranking}
One of the main applications of video popularity prediction is the proper ranking of the videos by their popularity. For instance, it allows the operating company to show the most popular videos on the main page, which always attracts a large share of user traffic\footnote{e.g., Bing shows ``most watched" videos at its main Video search page: http://www.bing.com/videos/browse}. Thus, we investigate how the quality measured in ranking metrics changes over predictions based on different feature sets.
In Table~\ref{NDCG100Table}, we present the average NDCG@100 over first 14 days for 7 main feature sets: {\tt API}, $\mathtt{ALL}\setminus\mathtt{WEB_{nag}}$, {\tt ALL}, {\tt LOG}, $\mathtt{WEB_{ag}}$, {\tt WEB}, and $\mathtt{WEB}\cup \mathtt{LOG}$. 

As one could see, the feature set {\tt ALL} clearly outperforms the  feature set  {\tt API} for all listed targets, and especially when we optimize prediction of non-transformed targets. Moreover, one could see that {\tt WEB} and $\mathtt{WEB}\cup \mathtt{LOG}$ notably outperform the baseline for cumulative targets, and $\mathtt{WEB}\cup \mathtt{LOG}$  slightly outperforms the baseline  or at least has the same value as the baseline for daily targets. \emph{Thus, one could conclude that for purposes of ranking tasks the Web and Log data could completely replace the API} data (if a company does not have its own logs, then still Web data outperforms API data for cumulative views). It is the answer to the {\bf RQ1} and {\bf RQ2}. That means that if the task of an operating company is to present the lists of the most popular videos for the OC's users, then: (a) it can be done independently from video service hostings, and (b) its costs on crawling and processing of embed and link data are justified.

\begin{table}
	\centering
 
		\caption{Comparison of API, Web and Log feature sets in terms of average NDCG@100 over first 14 days since video creation for targets optimized  by minimizing RMSE (in \% of NDCG@100 with respect to {\tt API} performance).\label{NDCG100Table}}{%

	\begin{tabular}{|l||cc|cc||cc|cc|} 
		\hline
		\multirow{2}{*}{Features} & \multicolumn{4}{|c||}{Cumulative targets ($\theta\in\mathbf{\Theta}$)}
				& \multicolumn{4}{|c|}{Daily targets ($\theta\in\mathbf{\Theta}$)} \\  
		\cline{2-9}
		& \multicolumn{2}{|c|}{{\tt log}-transformed} & \multicolumn{2}{|c||}{non-transformed} 
		& \multicolumn{2}{|c|}{{\tt log}-transformed} & \multicolumn{2}{|c|}{non-transformed}    \\ 
		\hline
		\hline
		{  \tt API	} & $ \underline{0.391} $	& $  0\%  $  & $ 0.334 $	 & $  0\%  $ &  $ \underline{0.37} $	& $  0\%  $ &$ 0.361 $ & $  0\%  $\\
		\hline
		{  $\mathtt{All}  \setminus  \mathtt{WEB_{nag}}$ }	& $ 0.466 $	 & $  \mathbf{+19.25\%}  $		 & $ \underline{0.506} $	 & $ +51.62\%  $	 & $ 0.398 $	 & $  +7.45\%  $		 & $ \underline{0.495} $	 & $ +37.13\%  $ \\
		\hline
		{  $\mathtt{All}$ }	& $ 0.426 $	 & $  +8.86\%  $		 & $ \underline{0.511} $	 & $ \mathbf{+53.06\%}  $	&  $ 0.398$	 & $  \mathbf{+7.5\%}  $ & $ \underline{0.499}$	 & $  \mathbf{+38.2\%}  $ \\
		\hline
		{  \tt LOG }	 & $ \underline{0.235}$	 & $  -39.86\%  $	  & $ 0.136$	 & $  -59.24\%  $ &  $ \underline{0.236}$	 & $  -36.16\%  $	 & $ 0.137$	 & $  -62.01\%  $	 \\
		\hline
		{  $\mathtt{WEB_{ag}}$ }	 & $ \underline{0.435} $	 & $ +11.20\%  $	  & $ 0.406$	 & $  +21.59\%  $	 & $ \underline{0.337} $	 & $ -8.89\%  $	  & $ 0.332$	 & $  -7.97\%  $ \\
		\hline
		{  \tt WEB }	 & $ \underline{0.442} $	 & $ +12.85\%  $	  & $ 0.405$	 & $  +21.28\%  $	 &  $ \underline{0.34}$	 & $  -8.16\%  $	 & $ 0.337$	 & $  -6.75\%  $ \\
		\hline
		{  $\mathtt{WEB}\cup \mathtt{LOG}$ }	& $ \underline{0.464}$	 & $  +18.48\%  $	 & $ 0.455$	 & $  +36.4\%  $	 & $ 0.37$	 & $  0\%  $	 & $ \underline{0.378}$	 & $  +4.74\%  $	\\
		\hline

	\end{tabular}
}
\end{table}

Table~\ref{NDCG100Table} shows that implementation of the non-aggregated web features has a solid profit: their usage improves the prediction quality of aggregated features with respect to both the Web features only ($\mathtt{WEB}$ vs $\mathtt{WEB_{ag}}$: by 1.65\% for cumulative, 0.73\%  for daily log-transformed views, and 1.22\%  for daily views  without transformation) and all features ($\mathtt{ALL}$ vs $\mathtt{ALL}\setminus\mathtt{WEB_{nag}}$: by 1.44\% for cumulative and 1.07\%  for daily views  without transformation).

From Table~\ref{NDCG100Table} one could also learn another lesson. Since the logarithmic transformation is monotonic, the original ranks of videos both in terms of the $log$-transformed and non-transformed target values are the same. Thus, the difference in average NDCG@100  values for  the $log$-transformed and non-transformed targets gives the answer to the question: Does the log-transformation of views in the optimization of RMSE improve the ranking results? We underlined the best result between them for each feature set, independently both for cumulative and for daily views. One can conclude that \emph{the transformation gives significant improvement for some, but not all sets of features.} The improvement is  more visible for  cumulative views than for the daily ones.

\section{Conclusions}
\label{sec_Conclusions}
We investigated the utility of newly proposed data sources (the embeds of and the links to videos publicly available on the Web  and internal logs of Yandex) for the task of video popularity prediction and compared them with the data provided by the video hosting via its API. We prepared more than $100$ features collected from all available data sources and to answer to a number  of  related research questions. We used both simple feature models, and more complicated feature models (the linear influence model that utilizes the features in non-aggregated form).

We found that the new data sources allow to improve the video popularity prediction quality, and they are able to compensate for the absence of API starting from the 8th day since the day of the video upload, if a company is interested in prediction of exact values of the current popularity. The new data could also compensate for any of the missing groups of API features that we considered in our study.

We also examined the case with a popular search engine, which predicts  popularity of videos in order to present them in a proper order to its users. In that case we compared the relative performance between feature sets in terms of the ranking measure NDCG@100. We found that the web and log data could replace and even outperform the API data.

As future work we can, first, extend the set of feature source groups by investigating the data obtained from the API of another content operating company. Second, we can train different predictors for different topical categories of videos. Third, we can also experiment with training not the regression function and optimizing for RMSE, but a classifier to predict if the video belongs to the set of the most or the least popular ones. Finally, it would be interesting to study how to use online user feedback received on a list of the most popular videos presented on the main page in order to correct the prediction in real-time.

% Bibliography
\bibliographystyle{apalike}
\bibliography{2016-arxiv-pvp}

\medskip

\end{document}